\documentclass[letterpaper]{article} 
\usepackage{aaai2027}
\usepackage[hyphens]{url}  
\usepackage{graphicx} 
\usepackage{natbib}  
\usepackage{caption} 
\usepackage{algorithm}
\usepackage{algorithmic}

\usepackage{amsmath, amssymb, amsthm, mathtools}
\usepackage[capitalize]{cleveref}
\usepackage{thmtools}
\usepackage{thm-restate}

\theoremstyle{plain}
\newtheorem{theorem}{Theorem}
\newtheorem{lemma}{Lemma}
\newtheorem{corollary}{Corollary}
\theoremstyle{definition}
\newtheorem{assumption}{Assumption}
\newtheorem{definition}{Definition}
\theoremstyle{remark}

\DeclareMathOperator*{\argmax}{\arg\!\max}

\ifdefined\val\else\DeclareMathOperator{\val}{val}\fi
\providecommand{\cS}{\mathcal{S}}
\providecommand{\cA}{\mathcal{A}}
\providecommand{\cP}{\mathcal{P}}
\providecommand{\shapley}{\mathcal{T}}
\providecommand{\reg}{\operatorname{reg}}
\providecommand{\norm}[1]{\left\lVert #1 \right\rVert}
\providecommand{\abs}[1]{\left|#1\right|}
\providecommand{\Racc}{\mathring{\mathbf R}}
\providecommand{\Raccpf}{\widetilde{\mathbf R}}

\providecommand{\refCoverage}{\cref{ass:depth-d-coverage}}
\providecommand{\refSearchBackup}{\cref{lem:search-to-backup}}
\providecommand{\refFittedDepthD}{\cref{thm:fitted-depth-d}}
\providecommand{\labelEndToEnd}{\label{thm:end-to-end-fitted-search}}
\providecommand{\labelEndToEndUniform}{\label{eq:end-to-end-uniform}}
\providecommand{\labelEndToEndMgBound}{\label{eq:end-to-end-mg-bound}}
\providecommand{\labelEndToEndFitBound}{\label{eq:end-to-end-fit-bound}}
\providecommand{\labelImperfectTransfer}{\label{thm:imperfect-regret-transfer}}
\providecommand{\labelWCondition}{\label{eq:w-condition}}
\providecommand{\labelApproxNash}{\label{lem:approx-nash}}

\newcommand{\BodyExactDepthD}{%
\label{thm:depth-d-sup-contraction}%
For every integer $D\ge0$ and all bounded $V,W$,
\[
    \norm{\shapley^D V-\shapley^D W}_\infty
    \le
    \gamma^D\norm{V-W}_\infty.
\]
In particular, because $\shapley^D V^*=V^*$, a frontier approximation $V$ obeys
\[
    \norm{\shapley^D V-V^*}_\infty
    \le
    \gamma^D\norm{V-V^*}_\infty.
\]%
}

\newcommand{\BodyFittedDepthD}{%
\label{thm:fitted-depth-d}%
Suppose \refCoverage{} holds and define
\[
    \rho=\gamma^D C_{p,\mu,D}.
\]
If $\rho<1$, then for every integer $K\ge1$,
\begin{align}
    \norm{V_K-V^*}_{p,\mu}
    \le{}&
    \rho^K\norm{V_0-V^*}_{p,\mu}\notag\\
    &+
    \sum_{j=0}^{K-1}
    \rho^{K-1-j}
    \left(
        \varepsilon_{\mathrm{mg},j}
        +
        \varepsilon_{\mathrm{fit},j}
    \right).
    \label{eq:fitted-finite-bound}
\end{align}
If the errors are uniformly bounded by
$\varepsilon_{\mathrm{mg},j}\le\varepsilon_{\mathrm{mg}}$ and
$\varepsilon_{\mathrm{fit},j}\le\varepsilon_{\mathrm{fit}}$, then
\begin{equation}\label{eq:fitted-uniform-bound}
    \norm{V_K-V^*}_{p,\mu}
    \le
    \rho^K\norm{V_0-V^*}_{p,\mu}
    +
    \frac{1-\rho^K}{1-\rho}
    \left(
        \varepsilon_{\mathrm{mg}}
        +
        \varepsilon_{\mathrm{fit}}
    \right).
\end{equation}
Consequently,
\[
    \limsup_{K\to\infty}\norm{V_K-V^*}_{p,\mu}
    \le
    \frac{\varepsilon_{\mathrm{mg}}+\varepsilon_{\mathrm{fit}}}{1-\rho}.
\]%
}

\newcommand{\BodyEndToEnd}{%
\labelEndToEnd%
Suppose the conditions of \refFittedDepthD{} and
\refSearchBackup{} hold for $k=0,\ldots,K-1$, and fix
$\eta>0$ and $\delta\in(0,1)$. For each $k$, use state-dependent budgets $t_k(s)$ satisfying \refSearchBackup{} and form the targets $Y_k$.
Conditional on the preceding training history and $Y_k$, suppose
$s_{k,1},\ldots,s_{k,n_k}$ are i.i.d.\ from $\mu$. Define
\begin{align*}
    \ell_k(V;s)
    &:=
    \abs{V(s)-Y_k(s)}^p,\\
    \widehat{\mathcal L}_{k,n_k}(V)
    &:=
    \frac1{n_k}\sum_{m=1}^{n_k}\ell_k(V;s_{k,m}),
\end{align*}
and suppose $\ell_k(V;s)$ lies in a sample-independent interval of width
$\Delta_{\ell,k}$ for every finitely describable candidate $V$. Assume
a bounded value regressor Kolmogorov complexity $\kappa(V_k)\le\bar\kappa$ for every $k=0,\ldots,K$. Suppose, for
every $k=0,\ldots,K-1$, that $n_k=n$,
$\widehat{\mathcal L}_{k,n}(V_{k+1})\le\overline{\mathcal L}$,
$\Delta_{\ell,k}\le\Delta_\ell$. Then, jointly over search and fitting randomness,
with probability at least $1-\delta$,
\begin{equation}
    \norm{V_K-V^*}_{p,\mu}
    \le
    \rho^K\norm{V_0-V^*}_{p,\mu}
    +
    \frac{1-\rho^K}{1-\rho}
    \left(
        \varepsilon_{\mathrm{mg}}
        +
        \varepsilon_{\mathrm{fit}}
    \right),
    \labelEndToEndUniform
\end{equation}
where, simultaneously for every $k=0,\ldots,K-1$,
\begin{align}
    \varepsilon_{\mathrm{mg},k}
    &\le
    \varepsilon_{\mathrm{mg}}
    :=
    \Delta_{\mathrm{tree}}
    \left(D\epsilon_{\mathrm{exp}}+\eta\right),
    \labelEndToEndMgBound\\
    \varepsilon_{\mathrm{fit},k}
    &\le
    \varepsilon_{\mathrm{fit}}
    :=
    \left[
        \overline{\mathcal L}
        +
        \Delta_\ell
        \sqrt{
            \frac{
                \bar\kappa\log2+\log(K/\delta)
            }{
                2n
            }
        }
    \right]^{1/p}.
    \labelEndToEndFitBound
\end{align}
}

\newcommand{\BodyRegretDefs}{%
Consider two games $\Gamma$ and $\Gamma'$ with identical action spaces and
utility functions $u_i$ and $u_i'$, respectively. For a strategy profile
$(\sigma_{i,t},\sigma_{-i,t})$, the immediate regret of player $i$ for action
$a_i\in\cA_i$ in $\Gamma$ is
\[
\reg_{i,t}(a_i) = u_i(a_i, \sigma_{-i,t}) - u_i(\sigma_{i,t}, \sigma_{-i,t}) \, .
\]
The immediate regret $\reg'_{i,t}(a_i)$ in $\Gamma'$ is defined analogously by
replacing $u_i$ with $u_i'$.
The cumulative regret is
$\mathbf R^c_{i,T}(a_i) = \sum_{t=1}^T \reg_{i,t}(a_i)$,
and the average regret is
$\bar{\mathbf R}_{i,T}(a_i) = \tfrac{1}{T}\mathbf R^c_{i,T}(a_i)$.
The regret-matching potential is defined on cumulative regrets:
\[
\Phi(\mathbf z) = \sum_{a_i\in\cA_i} z(a_i)_+^2.
\]%
}

\newcommand{\BodyImperfectTransfer}{%
\labelImperfectTransfer%
Fix player $i$. Let $\Delta_u$ be a common bound on the payoff range in both games, and
\[
\sup_{a_i, a_{-i}}|u_i'(a_i, a_{-i}) - u_i(a_i, a_{-i})| \le \delta_u \, .
\]
After $T_1$ regret-matching iterations in $\Gamma$, suppose the fitted average
regret satisfies
$\| \widehat{\bar{\mathbf R}}_{i,T_1} - \bar{\mathbf R}_{i,T_1}\|_\infty \le \delta_R$.
For $w\ge 0$, initialize regret matching in $\Gamma'$ with
$\Racc_{i,T_1}=wT_1\widehat{\bar{\mathbf R}}_{i,T_1}$ and run it for $T_2$ additional iterations.
Write $T=T_1+T_2$ and use $w$ such that
\begin{equation} \labelWCondition
\Phi(\Racc_{i,T_1}) \le wT_1|\cA_i|\Delta_u^2 .
\end{equation}
Let $\mathbf R^{c,\prime}_{i,T_1}$ denote the regret of the first $T_1$ strategy
profiles evaluated in $\Gamma'$, and define
$\mathbf S^{w,\prime}_{i,T}=w\mathbf R^{c,\prime}_{i,T_1}+\sum_{t=T_1+1}^{T}\reg_{i,t}'$.
Then
\begin{align*}
\mathcal R_i^{w,\prime}(T_1,T_2)
&\equiv
\frac{\max_{a_i}\mathbf S^{w,\prime}_{i,T}(a_i)}{wT_1 + T_2} \\
&\le
\frac{\Delta_u\sqrt{|\cA_i|}}{\sqrt{wT_1 + T_2}}
+
\frac{wT_1(2\delta_u +\delta_R)}{wT_1 + T_2}.
\end{align*}
}

\newcommand{\BodyTransferHannan}{%
\label{cor:transfer-hannan}%
Under the conditions of \cref{thm:imperfect-regret-transfer}, fix $T_1$ and $w$.
Then the regret-matching solver in $\Gamma'$ is Hannan consistent:
\[
    \limsup_{T_2\to\infty}\mathcal R_i^{w,\prime}(T_1,T_2)\le 0.
\]
Thus any fixed finite transferred initialization changes finite-time behavior
but not the solver's asymptotic external-regret guarantee.%
}

\newcommand{\BodyApproxNash}{%
\labelApproxNash%
Suppose $\Gamma'$ is a two-player zero-sum game and both players satisfy the
conditions of \cref{thm:imperfect-regret-transfer} with the same weighting
convention. Suppose further that for each player $i$ we have a fitted average
strategy $\widehat{\bar\sigma}_{i,T_1}$ satisfying
$\|\widehat{\bar\sigma}_{i,T_1}-\bar\sigma_{i,T_1}\|_\infty \le \delta_\sigma$,
with $\bar\sigma_{i,T_1}(a_i)=\tfrac{1}{T_1}\sum_{t=1}^{T_1}\sigma_{i,t}(a_i)$,
and define the fitted weighted average strategy for player $i$ by
\[
\widehat{\bar\sigma}_i^{w,\prime}(a)
=
\frac{
wT_1\widehat{\bar\sigma}_{i,T_1}(a_i)
+
\sum_{t=T_1+1}^{T}\sigma_{i,t}(a_i)
}{
wT_1+T_2
}.
\]
Then the fitted weighted average strategies form an $\epsilon_{\mathrm{NE}}$-Nash equilibrium with
\begin{align*}
\epsilon_{\mathrm{NE}}
\le
2&\max_i
\left[
\frac{\Delta_u\sqrt{|\cA_i|}}{\sqrt{wT_1 + T_2}}
+
\frac{wT_1(2\delta_u+\delta_R)}{wT_1 + T_2}
\right] \\
&+
\Delta_u
\frac{wT_1}{wT_1+T_2}
\left(|\cA_1|+|\cA_2|\right)\delta_\sigma,
\end{align*}
where $\Delta_u$ is the common payoff-range bound.%
}

\newcommand{\BodyAsymptoticTreeEq}{%
\label{cor:transfer-sm-mcts-a}%
Let $\mathcal G_D$ be a two-player zero-sum simultaneous-move game of depth $D$.
At every node, run the fixed-finite-transfer solver with uniform
$\epsilon_{\mathrm{exp}}$-exploration, where $\epsilon_{\mathrm{exp}}\in(0,1)$. Then, for every $\xi>0$,
there almost surely exists $t_0$ such that the empirical frequencies produced
by SM-MCTS-AP form a subgame-perfect
$(2D(D+1)\epsilon_{\mathrm{exp}}+\xi)$-equilibrium for all $t\ge t_0$.%
}

\crefname{algorithm}{Algorithm}{Algorithms}
\Crefname{algorithm}{Algorithm}{Algorithms}
\providecommand{\transfer}{\Psi}
\providecommand{\oracle}{\theta}
\providecommand{\Vnet}{V_{\oracle}}
\providecommand{\signet}{\bar\sigma_{\oracle}}
\providecommand{\Rnet}{\bar{\mathbf R}_{\oracle}}
\providecommand{\Unif}{\mathrm{Unif}}

\providecommand{\crefSimulate}{\cref{alg:simulate}}

\newcommand{\BodySearch}{%
\small
\begin{algorithmic}[1]
\REQUIRE root state $s$; oracle $\oracle=(\Vnet,\signet,\Rnet)$
\REQUIRE budget $T$; depth $D$; exploration $\epsilon_{\mathrm{exp}}$
\ENSURE  strategies $(\bar\sigma_1,\bar\sigma_2)$; regret targets $y_R$; root value $v$
\STATE $h_0\leftarrow$ \textsc{NewNode}$(s)$;\;
       $d(h_0)\leftarrow0$;\; $q(h_0)\leftarrow1$
\FOR{$t=1$ \TO $T$}
    \STATE \textsc{Simulate}$(h_0,\oracle,\epsilon_{\mathrm{exp}})$
           \COMMENT{\crefSimulate}
\ENDFOR
\FORALL{$i\in\{1,2\}$}
    \STATE $\bar\sigma_i\leftarrow
           \mathbf c_i(h_0)/\lVert\mathbf c_i(h_0)\rVert_1$
           \COMMENT{avg.\ strategy}
    \STATE $y_{R,i}\leftarrow
           \Raccpf_i(h_0)\big/
           \textstyle\sum_{\mathbf a}N(h_0,\mathbf a)$
           \COMMENT{avg.\ regret target}
\ENDFOR
\STATE $v\leftarrow G(h_0)/N(h_0)$
       \COMMENT{root average payoff}
\RETURN $(\bar\sigma_1,\bar\sigma_2,v,y_R)$
\end{algorithmic}%
}

\newcommand{\BodyTrainingMu}{%
\small
\begin{algorithmic}[1]
\REQUIRE game $f,r$; initial oracle $\oracle_0$; state distribution $\mu$
\REQUIRE outer iterations $K$; sample size $n$; budget $T$; depth $D$
\ENSURE  trained oracle $\oracle_K$
\FOR{$k=0$ \TO $K-1$}
    \STATE draw $s_{k,1},\ldots,s_{k,n}
           \sim \mu$
    \STATE $\mathcal D_k\leftarrow\emptyset$
    \FOR{$m=1$ \TO $n$}
        \STATE $(\bar\sigma,v,y_R)\leftarrow$
               \textsc{SM-MCTS-AP}$(s_{k,m},\oracle_k,T,D,
               \epsilon_{\mathrm{exp}},\emptyset)$
               \COMMENT{no transfer}
        \STATE $Y_k(s_{k,m})\leftarrow v$
               \COMMENT{$=(\widehat{\shapley}_{D,k}V_k)(s_{k,m})$}
        \STATE $y_{\sigma,i}\leftarrow\bar\sigma_i$
        \STATE $\mathcal D_k\leftarrow\mathcal D_k
               \cup\{(s_{k,m},Y_k,y_\sigma,y_R)\}$
    \ENDFOR
    \STATE $\oracle_{k+1}\leftarrow$
           \textsc{Fit}$(\oracle_k,\mathcal D_k)$
           \COMMENT{minimize $\widehat{\mathcal L}_{k,n}$}
\ENDFOR
\RETURN $\oracle_K$
\end{algorithmic}%
}

\newcommand{\BodySelfPlay}{%
\small
\begin{algorithmic}[1]
\REQUIRE game $f,r$; initial oracle $\oracle_0$; horizon cap $H$
\REQUIRE outer iterations $K$; episodes $M$; budget $T$; depth $D$
\REQUIRE exploration schedule $\epsilon(\cdot)$; EMA decay $\tau$
\ENSURE  trained oracle $\oracle_K$
\STATE $\bar\oracle_0\leftarrow\oracle_0$
       \COMMENT{EMA self-play oracle}
\FOR{$k=0$ \TO $K-1$}
    \STATE $\mathcal D_k\leftarrow\emptyset$;\quad
           $\epsilon_k\leftarrow\epsilon(k)$
    \FOR{$m=1$ \TO $M$}
        \STATE $s\sim b_0$;\quad $t\leftarrow0$
        \WHILE{$s$ not terminal \AND $t<H$}
            \STATE $(\bar\sigma,v,y_R)\leftarrow$
                   \textsc{SM-MCTS-AP}$(s,\bar\oracle_k,T,D,
                   \epsilon_k,\emptyset)$
            \STATE append $(s,v,\bar\sigma,y_R)$ to $\mathcal D_k$
                   \COMMENT{targets as in Alg.~2}
            \STATE $a^i\sim(1-\epsilon_k)\bar\sigma_i
                   +\epsilon_k\Unif(\cA_i)$
            \STATE $s\leftarrow f(s,a^1,a^2)$;\quad
                   $t\leftarrow t+1$
        \ENDWHILE
    \ENDFOR
    \STATE $\oracle_{k+1}\leftarrow$
           \textsc{Fit}$(\oracle_k,\mathcal D_k)$
    \STATE $\bar\oracle_{k+1}\leftarrow
           \tau\bar\oracle_k+(1-\tau)\oracle_{k+1}$
\ENDFOR
\RETURN $\oracle_K$
\end{algorithmic}%
}

\newcommand{\BodySimulate}{%
\small
\begin{algorithmic}[1]
\REQUIRE node $h$; oracle $\oracle$; exploration $\epsilon_{\mathrm{exp}}$
\IF{$s_h\in\mathcal Z$}
    \RETURN $0$
\ENDIF
\IF{$d(h)\ge D$}
    \STATE $G(h)\mathrel{+}=\Vnet(s_h)$;\quad $N(h)\mathrel{+}=1$
    \RETURN $\Vnet(s_h)$ \COMMENT{learned frontier}
\ENDIF
\IF{$h\notin\mathcal T$}
    \STATE \textsc{Expand}$(h)$
    \RETURN $\Vnet(s_h)$ \COMMENT{no update on this visit}
\ENDIF
\STATE $(\mathbf a,\sigma)\leftarrow$ \textsc{Select}$(h)$
\STATE $h'\leftarrow$ \textsc{Child}$(h,\mathbf a)$
       \COMMENT{$d(h')=d(h)+1$, $q(h')=q(h)\,\bar\sigma_{1,\oracle}(a^1)\,\bar\sigma_{2,\oracle}(a^2)$}
\STATE $v'\leftarrow$ \textsc{Simulate}$(h',\oracle,
       \epsilon_{\mathrm{exp}})$
\STATE $u\leftarrow r(h,\mathbf a)+\gamma v'$
\STATE \textsc{Update}$(h,\mathbf a,\sigma,u,v')$
\RETURN $G(h)/N(h)$
\end{algorithmic}%
}

\newcommand{\BodyExpand}{%
\small
\begin{algorithmic}[1]
\STATE enumerate $\cA_1\times\cA_2$; store $r(h,\mathbf a)$ and
       $\widehat V(h,\mathbf a)\leftarrow\Vnet\bigl(f(s_h,\mathbf a)\bigr)$
\STATE $\bigl(\Racc^0_1,\Racc^0_2,\mathbf c^0_1,\mathbf c^0_2,m\bigr)
       \leftarrow\transfer\bigl(h,\oracle\bigr)$
       \COMMENT{prior transfer}
\STATE $\Racc_i(h)\leftarrow\Racc^0_i$;\quad
       $\mathbf c_i(h)\leftarrow\mathbf c^0_i$;\quad
       $\Raccpf_i(h)\leftarrow\mathbf 0$
\STATE $N(h)\leftarrow m$;\quad
       $N(h,\mathbf a)\leftarrow
       m\,\bar\sigma_{1,\oracle}(a^1)\,\bar\sigma_{2,\oracle}(a^2)$
\STATE $G(h)\leftarrow m\,\Vnet(s_h)$;\quad
       $\mathcal T\leftarrow\mathcal T\cup\{h\}$
\end{algorithmic}%
}

\newcommand{\BodySelect}{%
\small
\begin{algorithmic}[1]
\STATE $\sigma_i\leftarrow$ \textsc{RegretMatch}$\bigl(\Racc_i(h)\bigr)$,\;
       $i\in\{1,2\}$
\STATE $\mu_i\leftarrow(1-\epsilon_{\mathrm{exp}})\sigma_i
       +\epsilon_{\mathrm{exp}}\Unif(\cA_i)$
\STATE $a^1\sim\mu_1$, $a^2\sim\mu_2$
\RETURN $\bigl((a^1,a^2),\sigma\bigr)$
\end{algorithmic}%
}

\newcommand{\BodyUpdate}{%
\small
\begin{algorithmic}[1]
\STATE $\widehat V(h,\mathbf a)\mathrel{+}=
       \bigl(v'-\widehat V(h,\mathbf a)\bigr)/\bigl(N(h,\mathbf a)+1\bigr)$
\STATE $N(h)\mathrel{+}=1$;\; $N(h,\mathbf a)\mathrel{+}=1$;\;
       $G(h)\mathrel{+}=u$
\STATE $Q\leftarrow r(h,\cdot,\cdot)+\gamma\widehat V(h,\cdot,\cdot)$
       \COMMENT{current matrix game}
\STATE $\Delta_1(b)\leftarrow Q(b,a^2)-u$ for $b\neq a^1$,\;
       $\Delta_1(a^1)\leftarrow0$
\STATE $\Delta_2(b)\leftarrow u-Q(a^1,b)$ for $b\neq a^2$,\;
       $\Delta_2(a^2)\leftarrow0$
\STATE $\Racc_i(h)\mathrel{+}=\Delta_i$;\quad
       $\Raccpf_i(h)\mathrel{+}=\Delta_i$
       \COMMENT{RM$^+$: $[\,\cdot\,]_+$}
\STATE $\mathbf c_i(h)\mathrel{+}=\sigma_i$
\end{algorithmic}%
}

\usepackage{newfloat}
\usepackage{listings}
\DeclareCaptionStyle{ruled}{labelfont=normalfont,labelsep=colon,strut=off} 
\floatstyle{ruled}
\newfloat{listing}{tb}{lst}{}
\floatname{listing}{Listing}

\usepackage{booktabs}

\definecolor{claudeorange}{HTML}{D97757}

\title{Simultaneous AlphaZero: Extending Tree Search to Markov Games}

\title{Simultaneous AlphaZero: Extending Tree Search to Markov Games}
\author {
    Tyler Becker \textsuperscript{\rm 1},
    Zachary Sunberg \textsuperscript{\rm 1}
}
\affiliations {
    \textsuperscript{\rm 1}University of Colorado Boulder\\
}

\DeclareUnicodeCharacter{2212}{-}
\begin{document}

\maketitle

\begin{abstract}
Many strategic planning problems require agents to act simultaneously, making turn-based search unsuitable and requiring a normal form game to be solved at each tree state.
We introduce \emph{Simultaneous AlphaZero}, a learning and search method for continuous-state, two-player zero-sum deterministic Markov games.
A learned value function bootstraps finite-depth regret-matching search, while approximate regret transfer uses learned regret and average-strategy functions to warm start the local solver at deployment.
We establish complementary guarantees for both components.
First, we show that even a shallow finite-depth solver contracts errors in its learned frontier values, and we bound how local solution and value-fitting errors propagate through repeated learning and search.
Second, imperfect regret and average-strategy transfer yields a finite-time approximate-equilibrium bound decomposing ordinary regret, game mismatch, regret-fitting error, and strategy-fitting error; any fixed finite warm start preserves the solver's asymptotic no-regret guarantee.
Experiments in continuous Dubins pursuit-evasion and satellite custody maintenance compare unguided, value-only, and fully transferred search in direct solver cross-play.
Across both domains, oracle-guided search generally improves performance over unguided search, demonstrating effective game-theoretic planning in large continuous state spaces.
\end{abstract}


\section{Introduction}
Every goal-seeking agent that acts in the world, from physical vehicles to bots in cyberspace, will interact with other intelligent agents pursuing their own goals.
Yet, for the sake of tractability, much of reinforcement learning is designed for either single-agent or turn-taking settings. While single-agent methods can be applied to multi-agent problems, such as by fixing the other agents to assumed policies, these policies tend to be fragile as the other agents may not play according to the policy prescribed for training.
This fragility can be more concretely viewed as exploitability: the incentive that players have to deviate from their prescribed policy. In the aforementioned single-agent approach, the other agents frequently have ample incentive to deviate, and in zero-sum settings, this incentive to deviate is precisely the utility these agents can take away from the trained agent.

\begin{figure}
    \centering
    \includegraphics[width=\linewidth]{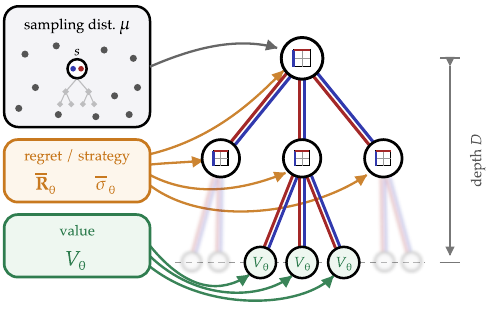}
    \caption{Simultaneous AlphaZero alternates between simultaneous-move tree search and fitting value, regret, and average-strategy priors}
    \label{fig:sm-az}
\end{figure}

In the same way that many real problems do not just have a single agent, many real-life problems are not strictly turn-based; agents reason and may take actions simultaneously, not necessarily waiting for opposing agents to finish their turn, as is frequently the case in the environments on which game-theoretic methods are benchmarked. In such circumstances, one could formulate the simultaneous interaction as turn-based, but this forces a solution that is too optimistic for the agent designed to go last as they are granted the ability to react to the opposing agent, an artifact of the formulation that may not be true to the real simultaneous environment. A further complication is that while deterministic or pure strategies are sufficient for solving perfect information turn-based games, simultaneous games necessitate mixed strategies to achieve unexploitable solutions.
The Markov game provides a suitable mathematical formulation for sequential simultaneous interaction, and the field of Multi-agent Reinforcement Learning (MARL) seeks to provide algorithms for solving Markov games~\cite{albrecht2024marl}.

The computation that enables intelligent agents to act can be divided into an \emph{offline} phase prior to deployment, and an \emph{online} phase immediately before an action is taken.
Popular reinforcement learning algorithms used for MARL such as Proximal Policy Optimization~\cite{yu2022mappo,rudolph2026reevaluating} typically train neural network policies offline and execute only minimal inference operations online.
Online methods such as Monte Carlo Tree Search (MCTS) \cite{browne2012mcts} have the advantage of focusing computation only on states reachable from the current state and can plan with any dynamic models or objectives defined at deployment time, but they are limited by deployment-time computation limits.
Our work builds on the AlphaZero \cite{silver2017alphazero} family of methods that combine neural network training and tree search both online and offline.

The original AlphaZero solved turn-taking games, and different versions have been applied to single agent Markov decision processes \cite{moerland2018a0c,shah2022nonasymptotic,schrittwieser2020muzero}. However, a complete AlphaZero-like approach for the more difficult simultaneous-play Markov Game setting has not been developed. Our work steps toward this goal by tackling an important subclass.

\paragraph{Contributions} We introduce \emph{Simultaneous AlphaZero} (SM-AZ), an algorithm for solving two-player zero-sum deterministic Markov games that combines both offline learning and online search. Its search component, SM-MCTS-AP (\cref{tab:algorithms}), is SM-MCTS-A~\cite{kovavrik2020hannan} augmented with learned priors: policy, regret, and value priors guide the tree search and the tree search yields improved policy, regret, and value targets that, in turn, improve the learned priors. We offer two primary theoretical guarantees: a finite-iteration error-propagation bound for fitted minimax value iteration, conditional on the error of each approximate tree-search backup and accompanied by a finite-sample regression bound. The second is a finite-time regret bound for approximate regret transfer in normal-form games that preserves the solver’s asymptotic no-regret guarantee. Finally, we offer empirical validation that demonstrates strong performance on a continuous pursuit-evasion scenario as well as a large satellite custody maintenance benchmark.

\section{Background}

The Markov Game (MG) is a mathematical formalism for problems where multiple agents make decisions sequentially to maximize an objective function~\cite{albrecht2024marl, kochenderfer2022dmu}.
A particular deterministic-transition MG instance is defined by the tuple $(\mathcal{N}, \mathcal{S}, \mathcal{A}, f, r, D, \gamma, b_0)$, where
$\mathcal{N}$ is the set of players ($i \in \mathcal{N}$) playing the game.
$\mathcal{S}$ is a set of possible states;
$\mathcal{A}$ is the Cartesian product of $\mathcal{A}_i$ where $\mathcal{A}_i$ is a set of player $i$'s actions;
$f$ is the deterministic state transition function where $s' = f(s,a)$ denotes the transition from state $s$ to state $s'$ via joint action $a$;
$r^i(s, a)$ is the scalar reward function for player $i$, given state $s$ and joint action $a$;
$D \in \mathbb{N}$ is the number of time steps in the horizon;
$\gamma \in [0,1)$ is the discount factor; and
$b_0$ is the initial distribution over states.

A policy for player $i$, $\pi^i$, is a mapping from state $s$, to a distribution over actions $\pi^i : \mathcal{S} \rightarrow \Delta(\mathcal{A}_i)$. The space of possible policies for agent $i$ is denoted $\Pi^i$.
A joint policy, $\pi = (\pi^1, \pi^2, \dots, \pi^{|\mathcal{N}|})$, is a collection of individual policies for each player.
The superscript $-i$ is used to mean ``all other players''. For example, $\pi^{-i}$ denotes the policies for all players except $i$.
Player $i$'s objective is to choose a policy to maximize his or her utility or value ($V$),
\begin{equation}
    V^i(\pi) = \mathbb{E}_{\pi}
    \left. \left[
        \sum_{t=0}^D \gamma^t r^i(s_t, a_t)
    \right\vert s_0 \sim b_0\right]\,.
\end{equation}
Since this objective depends on the joint policy, and the reward functions for individual players may not align with each other, MGs are not optimization problems where locally or globally optimal solutions are always well-defined.
Instead of optima, there are a variety of possible solution concepts.
The most common solution concept and the one adopted in this paper is the Nash equilibrium~\cite{nash1950equilibrium}.
A joint policy is a Nash equilibrium if every player is playing a best response to all others.
Mathematically a best response to $\pi^{-i}$ is a $\pi^i$ that satisfies
\begin{equation} \label{eq:br}
    V^i(\pi^i, \pi^{-i}) \ge V^i(\pi^{i\prime}, \pi^{-i})
\end{equation}
for all possible policies $\pi^{i\prime}$. In a Nash equilibrium, \cref{eq:br} is satisfied for all players. A relaxation of the Nash equilibrium is the $\epsilon$-Nash equilibrium, in which every player has an incentive to deviate that is \emph{at most} $\epsilon$:
\begin{equation}
    V^i(\pi^i, \pi^{-i}) \ge V^i(\pi^{i\prime}, \pi^{-i}) - \epsilon \,.
\end{equation}

A player $i$'s best response to all other players' policies $\pi^{-i}$ is a policy that greedily maximizes player $i$'s utility, holding $\pi^{-i}$ constant i.e.
\begin{equation}
\mathbf{BR}^{i}(\pi^{-i}) \in \argmax_{\pi^{i}} V^i(\pi^i, \pi^{-i}) \,.
\end{equation}

A useful scalar metric for a distance to Nash is then the sum of all deviation incentives, referred to as the \emph{Nash gap} or \textsc{NashConv}:
\begin{equation} \label{eq:nashconv}
    \textsc{NashConv}(\pi) = \sum_{i \in \mathcal{N}}\left[ V^i(\mathbf{BR}^{i}(\pi^{-i}), \pi^{-i}) - V^i(\pi) \right]\,.
\end{equation}

\textsc{NashConv}$(\pi)$ is a nonnegative quantity that, when zero, indicates that $\pi$ is a joint policy that constitutes a Nash equilibrium by virtue of none of the players having any incentive to unilaterally deviate.

One particularly important taxonomic feature for MGs is the relationship between the agents' reward functions. In \emph{cooperative} games, all agents have the same reward function, $r^i(s, a) = r^j(s, a)$ for all $i$ and $j$. In a two player \emph{zero-sum} game, the reward functions of the two players are directly opposed and add to zero, $r^1(s, a) = -r^2(s, a)$. When there are no restrictions on the reward function, the term \emph{general-sum} is used to contrast with cooperative, zero-sum, or other special classes. This work focuses specifically on the two-player zero-sum case.

\subsection{AlphaZero Framework}

The original AlphaZero\cite{silver2017alphazero} is an algorithm that learns the quality of states in a game, as well as a policy that maximizes this quality.
AlphaZero leverages MCTS as a policy improvement operator, where a network acts as a baseline policy, MCTS improves the policy, and this improved policy becomes the target for the network, thereby continuously improving the quality of the network policy.

\subsection{Simultaneous-Move Games}
This work focuses on the two-player zero-sum, \emph{simultaneous-action} case, where both players choose actions at each step without observing the other's choice beforehand.
This contrasts with turn-based models, where the player moving second in each step can react optimally, often biasing outcomes.
Simultaneous-action MGs more faithfully represent problems such as adversarial orbital maneuvering and avoid the artificial advantages of sequential play, but they also pose greater computational challenges due to the need to resolve joint action selection at every decision point.

\subsection{Normal Form Games and Regret Matching}
\label{sec:bg-regret-matching}

Fixing the value of every successor state reduces a single state of a simultaneous-move MG to a \emph{normal form game} (NFG): each player $i$ chooses an action $a_i\in\cA_i$ simultaneously, and Player~1 receives $u_1(a_1,a_2)$ while Player~2 receives $u_2=-u_1$. We write $\Delta_u$ for a bound on the range of $u_1$. A \emph{mixed strategy} $\sigma_i\in\Delta(\cA_i)$ is a distribution over actions, and $u_i$ extends to mixed strategies in expectation. Because an NFG generally admits no pure equilibrium, solving one requires reasoning over mixed strategies, and by the minimax theorem every finite two-player zero-sum NFG has a value attained by some mixed profile.

Rather than solving each NFG exactly, iterative no-regret methods approximate an equilibrium through repeated self-play. At iteration $t$ each player plays $\sigma_{i,t}$ and measures, for every action, how much better that action would have been against the opponent's realized strategy. This is the \emph{immediate regret}
\begin{equation}\label{eq:bg-immediate-regret}
    \reg_{i,t}(a_i)
    =
    u_i(a_i,\sigma_{-i,t})-u_i(\sigma_{i,t},\sigma_{-i,t}),
\end{equation}
whose running sum $\mathbf R^c_{i,T}(a_i)=\sum_{t=1}^{T}\reg_{i,t}(a_i)$ is the \emph{cumulative regret} and $\bar{\mathbf R}_{i,T}=\tfrac1T\mathbf R^c_{i,T}$ the \emph{average regret}.

\emph{Regret matching}~\cite{hart2000regret-matching} turns this bookkeeping into a strategy by playing each action in proportion to its accumulated positive regret,
\begin{equation}\label{eq:bg-regret-matching}
    \sigma_{i,t+1}(a_i)
    =
    \frac{\mathbf R^c_{i,t}(a_i)_+}{\sum_{a_i'\in\cA_i}\mathbf R^c_{i,t}(a_i')_+},
\end{equation}
defaulting to the uniform strategy when every cumulative regret is nonpositive. Its guarantee follows from tracking the potential $\Phi(\mathbf z)=\sum_{a_i}z(a_i)_+^2$ on cumulative regrets: each iteration increases $\Phi$ by at most $|\cA_i|\Delta_u^2$, so $\Phi(\mathbf R^c_{i,T})\le|\cA_i|\Delta_u^2T$ and therefore
\begin{equation}\label{eq:bg-no-regret}
    \max_{a_i}\bar{\mathbf R}_{i,T}(a_i)
    \le
    \frac{\Delta_u\sqrt{|\cA_i|}}{\sqrt{T}}.
\end{equation}
Average regret thus vanishes at rate $O(1/\sqrt T)$, making regret matching \emph{Hannan consistent}~\cite{hannan1957approximation}. In a two-player zero-sum game, if both players' average regret is at most $\epsilon$, their \emph{average} strategies $\bar\sigma_{i,T}=\tfrac1T\sum_t\sigma_{i,t}$ form a $2\epsilon$-Nash equilibrium, so \cref{eq:bg-no-regret} is simultaneously an equilibrium-error bound. Furthermore, Hannan consistency is precisely the property a per-node selection rule must satisfy for simultaneous-move MCTS to converge to a subgame-perfect approximate equilibrium~\cite{lisy2013convergence,kovavrik2020hannan}, which is why regret matching, rather than a bandit rule such as UCB, is the natural node solver in this setting.

\subsection{Regret Transfer}
\label{sec:bg-regret-transfer}

The bound in \cref{eq:bg-no-regret} is stated for a solver started from zero regret. In practice one often solves a sequence of \emph{related} games, and restarting discards all accumulated work. \emph{Regret transfer}~\cite{brown2014regret} addresses this: given cumulative regrets produced by $T_1$ iterations in a game $\Gamma$, and a new game $\Gamma'$ differing only in its utilities, the solver for $\Gamma'$ is initialized from the transferred regret vector rather than from zero. Scaling that vector by a weight $w\ge0$ before continuing lets the warm start be discounted. The accumulated regret then behaves like $wT_1$ virtual iterations, so after $T_2$ further iterations the effective denominator in \cref{eq:bg-no-regret} becomes $wT_1+T_2$ rather than $T_2$. When the transferred regrets are exact, the potential inequality that produced \cref{eq:bg-no-regret} still holds at initialization for any $w\le1$, so the no-regret guarantee survives the warm start while the solver begins from an already-informed strategy.

Transfer in this exact form assumes the accumulated regrets can be \emph{re-evaluated} under the new utilities: one must know what the regret of each previously played strategy would have been in $\Gamma'$. That is reasonable when $\Gamma$ and $\Gamma'$ differ in a known, structured way, and it is the assumption we relax in \cref{sec:approx-transfer}, where the transferred quantities are instead read approximately from learned function approximators.

\section{Related Work}

A large focus of game-theoretic work is directed towards turn-taking or extensive form games, where, at a given time, only one player is allowed to take an action.
This formulation is well-suited for a majority of table-top games such as Chess, Go, and Shogi.
However, there exists a large subset of games where this is not true.
For example, pursuit evasion games have both players acting simultaneously, as one player always being able to wait and react to another player is unrealistic.
This also applies to very simple games like rock paper scissors.
While it is not possible to translate simultaneous-action games into perfect information extensive form games, it is possible to express them as imperfect information extensive form games.
Here, rather than players acting simultaneously, an equivalent game can be constructed where players act sequentially, but neither player knows the other player's move.
While these formulations are general to any information not shared between agents, their generality precludes solution methods from exploiting the fact that a Markov game can be thought of as a sequence of NFGs~\cite{shapley1953stochastic}.

\Citet{perolat2015approximate} demonstrate strong guarantees for approximate dynamic programming for Markov games; however, their methodology is restricted to finite, discrete state spaces and does not include online policy refinement. \Citet{tak2014mcts,schaeffer2009comparing} offer methods for online tree search in Markov games, proposing the use of upper confidence trees in Markov games, but do not provide any guarantees. \Citet{bosansky2016mcts,kovavrik2020hannan} similarly propose tree search methods for Markov games while also providing asymptotic convergence guarantees that are leveraged in this work.
\Citet{shah2020alphazero} come close to our methodology, aiming to provide a principled algorithm mirroring AlphaZero~\cite{shah2020alphazero} with guarantees, but their proposed methodology only works for turn-taking Markov games, whereas we consider simultaneous actions. 
Finally, \citet{sokota2025stratego} provide a methodology for online refinement of a policy that extends to partially observable games, however it is limited to a single update step, precluding the possibility of continuously improving policy strength given further search.

\section{Simultaneous AlphaZero}
\label{sec:method}

\begin{table}[t]
\centering
\caption{Algorithm quick reference.}
\label{tab:algorithms}
\begin{tabular}{@{}lp{0.65\columnwidth}@{}}
\toprule
\textbf{Algorithm} & \textbf{Description} \\
\midrule
SM-MCTS-A
    & Simultaneous-move MCTS with average backups~\cite{kovavrik2020hannan}\\
SM-MCTS-AP
    & SM-MCTS-A with regret, policy, and value \emph{priors} \\
SM-AZ
    & Simultaneous-move AlphaZero with value oracle and regret transfer \\
SM-AZ-$\mu$
    & SM-AZ with a fixed state-sampling distribution (theoretical tool) \\
SM-AZ ($-\mathrm{RT}$)
    & SM-AZ with a value oracle only (no \emph{regret transfer}) \\
\bottomrule
\end{tabular}
\end{table}

\emph{Simultaneous AlphaZero} (SM-AZ) pairs a prior-guided simultaneous-move tree search with a fitted training loop that supplies the priors. Because a Markov game is a tree of NFGs, we build the method from the inside out, in three steps, each earning the guarantee that the next one consumes.

We begin at a \emph{single} NFG. \Cref{sec:approx-transfer} introduces \emph{approximate regret transfer}: a regret-matching solver warm-started from learned (and therefore imperfect) regret and average-strategy estimates, whose weighted average regret and equilibrium gap we bound in finite time. \Cref{sec:rt-search} then composes that solver at every node of a depth-$D$ tree, yielding \emph{regret transfer search} (SM-MCTS-AP) and lifting the per-node guarantee to an asymptotic equilibrium of the whole tree. Finally, \cref{sec:simaz-mu} closes the loop with \mbox{SM-AZ-$\mu$}, which uses that search as an approximate minimax backup, refits the oracle to its output, and repeats; here the guarantees become a contraction on value error and a single end-to-end bound on the fully trained algorithm. Note that we provide guarantees for approximate regret transfer, SM-MCTS-AP, and SM-AZ-$\mu$, but not for SM-AZ which we use as an empirically strong algorithm that is functionally identical to SM-AZ-$\mu$ except that it uses self-play to generate states instead of sampling them from a fixed distribution $\mu$.
\Cref{tab:algorithms} provides a quick reference for these algorithms, and full proofs of every statement below appear in the appendix.

\subsection{Approximate Regret Transfer} \label{sec:approx-transfer}
We begin by introducing SM-MCTS-AP, a variant of SM-MCTS-A \cite{kovavrik2020hannan}, but with tree values, regrets, and strategies initialized via informed \emph{priors}. The inspiration for these priors stems from existing work on regret transfer \cite{brown2014regret}. \citet{brown2014regret} argue that, when solving one game via regret matching, should the utilities of that game shift, one need not entirely throw away the first solution, and can instead warm-start the solution to the new game with the solution to the old one. This is immediately applicable to our problem as we aim to solve a series of small finite-depth Markov subgames for which the terminal payoffs shift every time we learn a new leaf node value oracle. Similarly, ideally we can use previously constructed solutions in the form of SM-MCTS trees to inform the construction of future trees.

However, regret transfer assumes that it is possible to perfectly read off the regrets in the new game for the strategies produced in the old game, but because we operate over generally large continuous space with many trees being constructed, we have to rely on an approximate regression over sampled value, regret, and policy returns, precluding the ability to read any of these quantities off exactly. Therefore, we introduce a theoretical framework for \emph{approximate regret transfer}, wherein we simply reuse the regrets from previous solutions without translating them into regrets for the new game.


In \cref{thm:imperfect-regret-transfer}, we demonstrate that when utilities of the game shift by at most $\delta_u$ and approximated average regret ($\widehat{\bar{\mathbf{R}}}$) differs from the true regret target ($\bar{\mathbf{R}}$) by at most $\delta_R$, the resulting weighted average regret decays as $O(1/\sqrt{T_2})$ where $T_2$ is the number of regret matching iterations run post-warm-start.

\begin{theorem}[Imperfect regret transfer]
\BodyImperfectTransfer
\end{theorem}

Using this sublinear regret result, we then show in \cref{lem:approx-nash} that this warm-started regret matching solver reaches an $\epsilon$-Nash solution ($\epsilon_{\mathrm{NE}}$) with $\epsilon_{\mathrm{NE}}$ decaying as $O(1/\sqrt{T_2})$ with an additional bias induced by the approximate strategy transfer with bounded error $\delta_\sigma$.

\begin{lemma}[Approximate Nash equilibrium after imperfect transfer]
\BodyApproxNash
\end{lemma}

This asymptotic convergence to Nash therefore satisfies the Hannan consistency~\cite{hannan1957approximation} criterion (\cref{cor:transfer-hannan}): a fixed finite warm start perturbs finite-time behavior but leaves the average regret vanishing. Adding uniform forced exploration $\epsilon_{\mathrm{exp}}$ further renders the approximate transfer regret matching solver $\epsilon_{\mathrm{exp}}$-Hannan consistent \emph{with guaranteed exploration}, so that every action at every node continues to be sampled indefinitely~\citep[Lemma~7]{kovavrik2020hannan}. That pair of properties is exactly what SM-MCTS-A requires of a node NFG solver, and the approximate transfer solver therefore qualifies despite being warm-started from approximate quantities.

\subsection{Regret Transfer Search}
\label{sec:rt-search}

The approximate regret transfer guarantees, however, all concern a single NFG in isolation: \cref{lem:approx-nash} bounds the equilibrium gap at one node, and \cref{cor:transfer-hannan} controls that node's regret as its iteration count grows. SM-MCTS-A instantiates one such solver at \emph{every} node of the depth-$D$ tree and averages their payoffs upward, and it is this composition, licensed by the convergence analysis of \citet{kovavrik2020hannan}, that lifts a per-node guarantee to a statement about the tree as a whole. Using the approximate regret transfer solver as the node NFG solver in SM-MCTS-A then yields SM-MCTS-AP, for which we can guarantee asymptotic Nash convergence with the use of approximate value, regret, and strategy priors by inheriting the results from \citet{kovavrik2020hannan}.

\begin{algorithm}[t]
\caption{SM-MCTS-AP simulation}
\label{alg:simulate}
\BodySimulate
\end{algorithm}

\cref{alg:simulate} gives the recursion that SM-MCTS-AP runs from the root, once per iteration, for a budget of $T$ iterations; the root procedure simply repeats it and then reports each player's normalized cumulative strategy $\mathbf c_i(h_0)/\lVert\mathbf c_i(h_0)\rVert_1$, prior-free average regret target $\widetilde{\mathbf R}_i(h_0)/\sum_{\mathbf a}N(h_0,\mathbf a)$, and the root average $G(h_0)/N(h_0)$. A single simulation descends the tree by sampling a joint action at each node from the current regret-matching strategy mixed with uniform $\epsilon_{\mathrm{exp}}$-exploration, stopping at a terminal state or at depth $D$, where the learned value $\Vnet(s_h)$ supplies the frontier evaluation in place of a rollout. The first visit to a node instead \emph{expands} it, transferring discounted value, policy, and regret priors. On the way back up, each node forms the payoff $u=r(h,\mathbf a)+\gamma v'$, updates the entry of its local matrix game, and accumulates the resulting regrets and strategy before returning its running average $G(h)/N(h)$ to its parent. The node-local subroutines (\textsc{Expand}, \textsc{Select}, and \textsc{Update}) are given in \cref{alg:subroutines}.

\subsection{\mbox{SM-AZ-$\mu$}: Learning the Priors}
\label{sec:simaz-mu}
Everything so far has taken the value oracle as given: SM-MCTS-AP solves a finite-depth subgame \emph{provided} value, policy, and regret priors. What supplies that oracle is the remaining piece, and it is supplied by the search itself: the root value of \cref{sec:rt-search} is an approximate minimax backup, so fitting to it constitutes a value-iteration step.

\paragraph{Training.}
\cref{alg:training} defines \mbox{SM-AZ-$\mu$}, the training procedure our analysis covers. At every outer iteration it draws a fresh i.i.d.\ sample $s_{k,1},\ldots,s_{k,n}\sim\mu$ from a \emph{fixed} state distribution, runs SM-MCTS-AP at each sampled state, and refits the oracle to the resulting targets. This is fitted depth-$D$ minimax value iteration: the target $Y_k(s)$ is the root average payoff, and the refit minimizes the empirical loss whose generalization error is controlled in \cref{thm:end-to-end-fitted-search}. 


\begin{algorithm}[t]
\caption{\mbox{SM-AZ-$\mu$}}
\label{alg:training}
\BodyTrainingMu
\end{algorithm}

\paragraph{Value convergence.}

We demonstrate convergence of SM-AZ-$\mu$ in the $L_p$-norm ($\|\cdot\|_{p,\mu}$) rather than the supremum norm ($\|\cdot\|_\infty$), as the supremum-norm bounds become overly conservative, even vacuous, when applied to sampling and fitting.

For a probability measure $\mu$ on $\cS$ and $p\in[1,\infty)$, define
\[
\norm{g}_{p,\mu}:=\left(\int_{\cS}\abs{g(s)}^p\,d\mu(s)\right)^{1/p}.
\]

By leveraging results from existing work using the $L_p$ norm for contraction and error propagation bounds for approximate dynamic programming~\cite{perolat2015approximate, munos2008fvi}, we are able to establish a convergence bound for SM-AZ-$\mu$ in terms of the quality of the value function fit $(\varepsilon_{\mathrm{fit},k} \equiv \norm{V_{k+1}-\widehat{\shapley}_{D,k}V_k}_{p,\mu})$ on iteration $k$ and the quality of the tree search solution approximating the depth-$D$ backup $(\varepsilon_{\mathrm{mg},k} \equiv \norm{\widehat{\shapley}_{D,k}V_k-\shapley^D V_k}_{p,\mu})$. Combining this convergence bound (\cref{thm:fitted-depth-d}) with Nash value estimate bounds for SM-MCTS-AP as well as empirical to true loss relation established in \citet{wilson2025mysterious}, we arrive at a conditional end-to-end value-error bound for SM-AZ-$\mu$ convergence (\cref{thm:end-to-end-fitted-search}).

\begin{theorem}[End-to-end \mbox{SM-AZ-$\mu$} value bound]
\BodyEndToEnd
\end{theorem}

\section{Experimental results}

We evaluate on two continuous-state, two-player zero-sum domains. The first is a Dubins Tag, in which an attacker attempts to reach a goal region while a defender attempts to intercept it; it serves as a controlled setting in which the transfer theory can be tested directly. Both agents are constrained to move according to Dubins' car dynamics ($\dot{x} = V\cos(\theta), \dot{y} = V\sin(\theta), \dot{\theta} = u$), where each agent has control over steering angle $u$, which is discretized into $u \in \{-45^{\circ}, 0^{\circ}, +45^{\circ}\}$. The second is a larger space domain awareness (SDA) custody-maintenance scenario in low Earth orbit, in which an observer satellite must maintain line-of-sight custody of a maneuvering target under realistic orbital mechanics and sensing constraints. Both agents are constrained to move in a 2-dimensional equatorial orbital plane and are granted the ability to perform in-plane impulse burns at every step, incrementing velocity with $\Delta v \in \{-100 \text{m/s}, 0 \text{m/s},+100 \text{m/s}\}$. The observer is equipped with a photometric sensor with a signal-to-noise (SNR) model outlined in \citet{coder2016snr}. The reward received by the observer satellite is precisely the SNR of the photometric sensor viewing the target satellite. The target satellite has the ability to break line-of-sight by occluding itself behind the Earth or may even be in line-of-sight of the observer but render the photometric sensor useless by hiding in the shadow cast by the Earth. 

\subsection{SM-MCTS-AP}
\begin{figure}
  \centering
  \includegraphics[width=0.8\columnwidth]{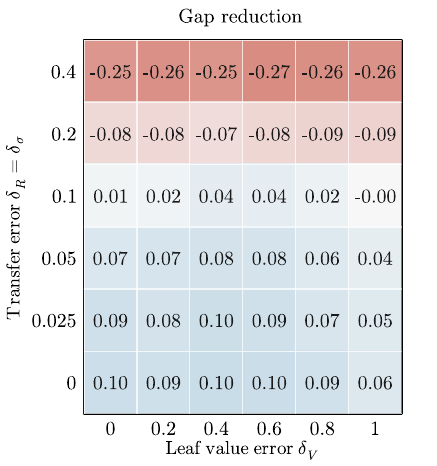}
  \caption{Reduction in the local-search Nash gap from regret and strategy
  transfer. The horizontal axis is the uniform leaf-value error bound
  $\delta_V$, and the vertical axis is the shared regret- and strategy-transfer
  error bound $\delta_R=\delta_\sigma$.  Each annotated cell reports the
  32-trial mean paired difference
  $\operatorname{NashGap}(\text{SM-MCTS-A})-
  \operatorname{NashGap}(\text{SM-MCTS-AP})$ after $T_2=64$ true-game updates;
  blue (positive) favors transfer and red (negative) favors restarting.
  Transferred priors come from $T_1=64$ source-game updates. }
  \label{fig:transfer-heatmap}
\end{figure}

We demonstrate the effectiveness of SM-MCTS-AP against SM-MCTS-A in \cref{fig:transfer-heatmap} on an isolated depth-five instance of the Dubins tag environment. We pair the true game $\Gamma'$, which emits the correct terminal payoffs, with a source game $\Gamma$ whose terminal payoffs are perturbed uniformly within $[-\delta_V,\delta_V]$.  At every nonterminal node, the source solver performs $T_1=64$ full-remaining-depth updates; its average regrets and strategies are then transferred with errors bounded by $\delta_R$ and $\delta_\sigma$.  Both the warm-started SM-MCTS-AP solver and the restarting SM-MCTS-A baseline subsequently receive $T_2=64$ full-remaining-depth updates using the true terminal payoffs. Across 32 paired trials, exact transfer reduces the mean Nash gap from $0.251$ to $0.150$, a $40\%$ reduction. At $\delta_V=0$, the mean advantage remains positive through $\delta_R=\delta_\sigma=0.1$ and reverses by $0.2$, indicating that error in regret transfer dominates the effect of leaf value error. 

\subsection{SM-AZ}
Next, we benchmark three algorithms in direct cross-play: SM-MCTS-A, SM-AZ (\(-\mathrm{RT}\)), and SM-AZ, as outlined in \cref{tab:algorithms}.

For direct solver cross-play, let $\widehat u(A_1,B_2)$ denote the empirical Player~1 utility when solver $A$ controls Player~1 and solver $B$ controls Player~2. We report the seat-balanced utility of $A$ against $B$ as
    $\widehat u_{\mathrm{sb}}(A,B)
    = \frac{1}{2}\left[
        \widehat u(A_1,B_2)-\widehat u(B_1,A_2)
      \right]\,.$
Thus, each solver is weighted equally in both player assignments, and a positive value favors the row solver. Seat-balanced standard errors use the corresponding independent-error propagation across the two ordered matchups. As the utilities for these games are asymmetric, the seat-balanced utility provides a better metric for advantage over another solver that accounts for the inherent asymmetry of a given game. 

\begin{table}[t]
\centering
\caption{Direct solver cross-play for Dubins Tag with 50 tree queries. The regret-transfer solver uses transfer weight $w=0.025$. Entries report mean utility $\pm$ standard error over 1,000 rollouts per ordered matchup.}
\label{tab:dubin-solver-round-robin}
\begingroup
\small
\setlength{\tabcolsep}{4pt}
\begin{tabular}{@{}lrrr@{}}
\toprule
\multicolumn{4}{c}{Player 1 utility: row solver as Player 1} \\
\cmidrule(lr){1-4}
Row $\backslash$ P2 solver
    & SM-MCTS-A & SM-AZ (−RT) & SM-AZ  \\
\midrule
SM-MCTS-A  & $ 0.03 \pm 0.01$ & $-0.59 \pm 0.01$ & $-0.61 \pm 0.01$ \\
SM-AZ (-RT) & $ 0.62 \pm 0.01$ & $-0.29 \pm 0.02$ & $-0.39 \pm 0.02$ \\
SM-AZ   & $ 0.59 \pm 0.01$ & $-0.28 \pm 0.02$ & $-0.40 \pm 0.02$ \\
\midrule
\multicolumn{4}{c}{Seat-balanced utility of the row solver} \\
\cmidrule(lr){1-4}
Row $\backslash$ Opponent
    & SM-MCTS-A & SM-AZ (-RT) & SM-AZ  \\
\midrule
SM-MCTS-A  & ---                  & $-0.60 \pm 0.01$ & $-0.60 \pm 0.01$ \\
SM-AZ (−RT) & $ 0.60 \pm 0.01$     & ---                  & $-0.05 \pm 0.01$ \\
SM-AZ   & $ 0.60 \pm 0.01$     & $0.05 \pm 0.01$ & ---                  \\
\bottomrule
\end{tabular}
\endgroup
\end{table}

\begin{table}[t]
\centering
\caption{Direct solver cross-play for SDA with 50 tree queries. The regret-transfer solver uses transfer weight $w=0.025$. Entries report mean utility $\pm$ standard error over 1,000 rollouts per ordered matchup from the restricted LEO initial-state distribution.}
\label{tab:sda-solver-round-robin}
\begingroup
\small
\setlength{\tabcolsep}{4pt}
\begin{tabular}{@{}lrrr@{}}
\toprule
\multicolumn{4}{c}{Player 1 utility: row solver as Player 1} \\
\cmidrule(lr){1-4}
Row $\backslash$ P2 solver
    & SM-MCTS-A & SM-AZ (−RT) & SM-AZ \\
\midrule
SM-MCTS-A  & $14.63 \pm 0.57$ & $8.47 \pm 0.37$ & $8.41 \pm 0.35$ \\
SM-AZ (−RT) & $11.42 \pm 0.58$ & $7.29 \pm 0.42$ & $7.37 \pm 0.42$ \\
SM-AZ  & $13.97 \pm 0.59$ & $8.82 \pm 0.44$ & $9.36 \pm 0.44$ \\
\midrule
\multicolumn{4}{c}{Seat-balanced utility of the row solver} \\
\cmidrule(lr){1-4}
Row $\backslash$ Opponent
    & SM-MCTS-A & SM-AZ (−RT) & SM-AZ \\
\midrule
SM-MCTS-A  & ---                  & $-1.48 \pm 0.34$ & $-2.78 \pm 0.34$ \\
SM-AZ (−RT) & $ 1.48 \pm 0.34$     & ---                  & $-0.73 \pm 0.30$ \\
SM-AZ  & $ 2.78 \pm 0.34$     & $0.73 \pm 0.30$ & ---                  \\
\bottomrule
\end{tabular}
\endgroup
\end{table}

From \cref{tab:dubin-solver-round-robin,tab:sda-solver-round-robin}, we find that the value-only solver (without regret transfer) consistently outperforms SM-MCTS-A (zero oracle) and the full regret transfer solver outperforms both the value-only solver and SM-MCTS-A. However, while we do demonstrate that this regret transfer does confer an advantage, we note that, due to the sensitivity to regret transfer error found in \cref{fig:transfer-heatmap}, this advantage is only evident when using small transfer weights (for example $w=0.025$), and using larger transfer weights leads to an over-reliance on coarse approximate priors and consequently degrades solution quality.


\section{Conclusion}
This paper defines Simultaneous AlphaZero (SM-AZ), which combines simultaneous-move tree search with learned value, regret, and average-strategy priors for deterministic two-player zero-sum Markov games. We establish a finite-time bound for approximate regret transfer and a conditional fitted-value error propagation bound for SM-AZ-$\mu$ that is nonasymptotic in fitting iterations and regression samples. We demonstrate that regret transfer improves both local search and direct cross-play beyond value guidance alone. However, the theory assumes fixed-distribution sampling and asymptotic search, and the experiments use single trained models.

\section*{AI Use Disclosure}
AI-assisted tools were used for language editing, proof checking, and figure preparation. The authors verified all outputs and take full responsibility for the final content.

\bibliography{aaai2027}

\onecolumn
\appendix
\setcounter{algorithm}{0}
\renewcommand{\thealgorithm}{A\arabic{algorithm}}

\renewcommand{\labelEndToEnd}{\label{app-thm:end-to-end-fitted-search}}
\renewcommand{\labelEndToEndUniform}{\label{app-eq:end-to-end-uniform}}
\renewcommand{\labelEndToEndMgBound}{\label{app-eq:end-to-end-mg-bound}}
\renewcommand{\labelEndToEndFitBound}{\label{app-eq:end-to-end-fit-bound}}
\renewcommand{\labelImperfectTransfer}{\label{app-thm:imperfect-regret-transfer}}
\renewcommand{\labelWCondition}{\label{app-eq:w-condition}}
\renewcommand{\labelApproxNash}{\label{app-lem:approx-nash}}

\section{Theoretical Results}

We consider an infinite-horizon, discounted, two-player zero-sum Markov game with state space $\cS$, finite action spaces $\cA_1$ and $\cA_2$, bounded reward $r$, deterministic transition map $f:\cS\times\cA_1\times\cA_2\to\cS$, and discount $\gamma\in[0,1)$. Player 1 maximizes and Player 2 minimizes. For a bounded value function $V$, define the one-step payoff matrix
\begin{equation}
    Q_V(s,a^1,a^2)
    =
    r(s,a^1,a^2)+\gamma V\bigl(f(s,a^1,a^2)\bigr)
\end{equation}
and the exact Shapley operator
\begin{equation}\label{eq:shapley-operator}
    (\shapley V)(s)
    =
    \val\bigl(Q_V(s)\bigr)
    :=
    \max_{x\in\Delta(\cA_1)}
    \min_{y\in\Delta(\cA_2)}
    x^\top Q_V(s)y.
\end{equation}
The optimal value $V^*$ is the unique bounded fixed point of $\shapley$. Throughout, $D$ denotes the depth of the inner planning calculation, not the horizon of the underlying infinite-horizon game. We write $\shapley^0$ for the identity and $\shapley^D$ for the $D$-fold composition of the exact Shapley operator. For a probability measure $\mu$ on $\cS$ and $p\in[1,\infty)$, define
\[
    \norm{g}_{p,\mu}
    :=
    \left(\int_{\cS}\abs{g(s)}^p\,d\mu(s)\right)^{1/p}.
\]

\subsection{Exact depth-$D$ minimax contraction}

We first state the matrix-game fact used by every subsequent result.

\begin{lemma}[Lipschitz continuity of a matrix-game value]
\label{lem:matrix-value-lipschitz}
For any two payoff matrices $A$ and $B$ of the same dimensions,
\[
    \abs{\val(A)-\val(B)}
    \le
    \norm{A-B}_\infty,
\]
where $\norm{A-B}_\infty=\max_{i,j}\abs{A_{ij}-B_{ij}}$.
\end{lemma}

\begin{proof}
Let $\delta=\norm{A-B}_\infty$. For every pair of mixed actions $x,y$,
\begin{align*}
    x^\top(A-B)y
    &=
    \sum_{i,j}x_i y_j(A_{ij}-B_{ij})\\
    &\le
    \sum_{i,j}x_i y_j\abs{A_{ij}-B_{ij}}\\
    &\le
    \delta\sum_{i,j}x_i y_j\\
    &=\delta,
\end{align*}
where the last equality uses that $x$ and $y$ are probability vectors. Thus $x^\top Ay\le x^\top By+\delta$ for every $x,y$. Holding $x$ fixed and taking the minimum over $y$ preserves the inequality:
\[
    \min_y x^\top Ay
    \le
    \min_y x^\top By+\delta.
\]
Taking the maximum over $x$ then gives
\[
    \val(A)
    =\max_x\min_y x^\top Ay
    \le
    \max_x\min_y x^\top By+\delta
    =\val(B)+\delta.
\]
Interchanging $A$ and $B$ gives $\val(B)\le\val(A)+\delta$. Combining the two one-sided inequalities proves $\abs{\val(A)-\val(B)}\le\delta$.
\end{proof}

\begin{lemma}[Pointwise Shapley contraction]
\label{lem:pointwise-one-step}
For any bounded $V,W$ and any $s\in\cS$,
\[
    \abs{(\shapley V)(s)-(\shapley W)(s)}
    \le
    \gamma
    \max_{a^1,a^2}
    \abs{V\bigl(f(s,a^1,a^2)\bigr)
         -W\bigl(f(s,a^1,a^2)\bigr)}.
\]
Consequently,
\[
    \norm{\shapley V-\shapley W}_\infty
    \le
    \gamma\norm{V-W}_\infty.
\]
\end{lemma}

\begin{proof}
By definition of the payoff matrices, for each joint action,
\begin{align*}
    Q_V(s,a^1,a^2)-Q_W(s,a^1,a^2)
    &=r(s,a^1,a^2)+\gamma V\bigl(f(s,a^1,a^2)\bigr)\\
    &\quad-r(s,a^1,a^2)-\gamma W\bigl(f(s,a^1,a^2)\bigr)\\
    &=\gamma\left(
        V\bigl(f(s,a^1,a^2)\bigr)
        -W\bigl(f(s,a^1,a^2)\bigr)
    \right).
\end{align*}
Therefore, by \cref{eq:shapley-operator,lem:matrix-value-lipschitz},
\begin{align*}
    \abs{(\shapley V)(s)-(\shapley W)(s)}
    &=\abs{\val(Q_V(s))-\val(Q_W(s))}\\
    &\le \norm{Q_V(s)-Q_W(s)}_\infty\\
    &=\gamma\max_{a^1,a^2}
      \abs{V\bigl(f(s,a^1,a^2)\bigr)
           -W\bigl(f(s,a^1,a^2)\bigr)}.
\end{align*}
Moreover, every successor-state error is at most $\norm{V-W}_\infty$, so
\[
    \abs{(\shapley V)(s)-(\shapley W)(s)}
    \le\gamma\norm{V-W}_\infty
\]
for every $s$. Taking the supremum over $s$ proves the second statement.
\end{proof}

To retain the statewise information in this bound, let
\[
    \cP_D=(\cA_1\times\cA_2)^D
\]
be the set of length-$D$ joint-action paths. For the unique empty path, set $F^\emptyset(s)=s$. For $\tau=((a^1_0,a^2_0),\ldots,(a^1_{D-1},a^2_{D-1}))\in\cP_D$, let $F^\tau(s)$ be the state reached from $s$ after applying those joint actions in order under $f$.

\begin{lemma}[Pointwise depth-$D$ contraction]
\label{lem:pointwise-depth}
For every integer $D\ge0$, all bounded $V,W$, and every $s\in\cS$,
\[
    \abs{(\shapley^D V)(s)-(\shapley^D W)(s)}
    \le
    \gamma^D
    \max_{\tau\in\cP_D}
    \abs{V\bigl(F^\tau(s)\bigr)-W\bigl(F^\tau(s)\bigr)}.
\]
\end{lemma}

\begin{proof}
We proceed by induction on $D$. For $D=0$, $\shapley^0$ is the identity, $\cP_0=\{\emptyset\}$, $F^\emptyset(s)=s$, and $\gamma^0=1$. Hence
\begin{align*}
    \abs{(\shapley^0V)(s)-(\shapley^0W)(s)}
    &=\abs{V(s)-W(s)}\\
    &=\gamma^0\max_{\tau\in\cP_0}
      \abs{V(F^\tau(s))-W(F^\tau(s))}.
\end{align*}

Now suppose the claim holds at depth $D$. Applying \cref{lem:pointwise-one-step} to the bounded functions $\shapley^D V$ and $\shapley^D W$ gives
\begin{align*}
    \abs{(\shapley^{D+1}V)(s)-(\shapley^{D+1}W)(s)}
    &\le
    \gamma\max_{a^1,a^2}
    \abs{(\shapley^D V)\bigl(f(s,a^1,a^2)\bigr)
         -(\shapley^D W)\bigl(f(s,a^1,a^2)\bigr)}
\end{align*}
For each first joint action $(a^1,a^2)$, the induction hypothesis at the successor state $f(s,a^1,a^2)$ gives
\begin{align*}
    &\abs{(\shapley^D V)\bigl(f(s,a^1,a^2)\bigr)
         -(\shapley^D W)\bigl(f(s,a^1,a^2)\bigr)}\\
    &\qquad\le
    \gamma^D\max_{\tau\in\cP_D}
    \abs{V\bigl(F^\tau(f(s,a^1,a^2))\bigr)
         -W\bigl(F^\tau(f(s,a^1,a^2))\bigr)}.
\end{align*}
Substituting this inequality into the one-step bound yields
\begin{align*}
    \abs{(\shapley^{D+1}V)(s)-(\shapley^{D+1}W)(s)}
    &\le
    \gamma^{D+1}\max_{a^1,a^2}\max_{\tau\in\cP_D}
    \abs{V\bigl(F^\tau(f(s,a^1,a^2))\bigr)
         -W\bigl(F^\tau(f(s,a^1,a^2))\bigr)}.
\end{align*}
Every pair consisting of a first joint action $(a^1,a^2)$ and a tail path $\tau\in\cP_D$ determines a unique path $\tau'\in\cP_{D+1}$, and every $\tau'\in\cP_{D+1}$ has a unique such decomposition. For corresponding paths,
\[
    F^{\tau'}(s)=F^\tau(f(s,a^1,a^2)).
\]
Consequently,
\begin{align*}
    &\max_{a^1,a^2}\max_{\tau\in\cP_D}
    \abs{V\bigl(F^\tau(f(s,a^1,a^2))\bigr)
         -W\bigl(F^\tau(f(s,a^1,a^2))\bigr)}\\
    &\qquad=
    \max_{\tau'\in\cP_{D+1}}
    \abs{V\bigl(F^{\tau'}(s)\bigr)-W\bigl(F^{\tau'}(s)\bigr)}.
\end{align*}
This proves the depth-$(D+1)$ statement, so induction completes the proof.
\end{proof}

\begin{theorem}[Exact depth-$D$ minimax backup]
\BodyExactDepthD
\end{theorem}

\begin{proof}
For each $s\in\cS$, \cref{lem:pointwise-depth} gives
\begin{align*}
    \abs{(\shapley^D V)(s)-(\shapley^D W)(s)}
    &\le
    \gamma^D\max_{\tau\in\cP_D}
    \abs{V(F^\tau(s))-W(F^\tau(s))}\\
    &\le
    \gamma^D\norm{V-W}_\infty.
\end{align*}
Taking the supremum over $s$ on the left-hand side yields
\[
    \norm{\shapley^D V-\shapley^D W}_\infty
    \le
    \gamma^D\norm{V-W}_\infty.
\]

It remains to connect this contraction to frontier error relative to $V^*$. The identity $\shapley^D V^*=V^*$ follows by induction: $\shapley^0V^*=V^*$, and if $\shapley^dV^*=V^*$, then
\[
    \shapley^{d+1}V^*=\shapley(\shapley^dV^*)=\shapley(V^*)=V^*.
\]
Applying the first claim with $W=V^*$ therefore gives
\begin{align*}
    \norm{\shapley^D V-V^*}_\infty
    &=\norm{\shapley^D V-\shapley^D V^*}_\infty\\
    &\le\gamma^D\norm{V-V^*}_\infty,
\end{align*}
as claimed.
\end{proof}

\subsection{Fitted depth-$D$ minimax value iteration}

The supremum-norm result controls worst-case error without a coverage assumption. To state a fitted result in $L_p(\mu)$, we must additionally control how depth-$D$ search moves the fitting distribution $\mu$ through the state space.

\begin{assumption}[Depth-$D$ coverage]
\label{ass:depth-d-coverage}
Let $\mu$ be a probability measure on $\cS$ and fix $p\in[1,\infty)$. For every $\tau\in\cP_D$, assume the pushforward measure $F^\tau_*(\mu)$ is absolutely continuous with respect to $\mu$, and let
\[
    h_\tau
    =
    \frac{d(F^\tau_*(\mu))}{d\mu}.
\]
Here $F^\tau_*(\mu)$ denotes the pushforward of $\mu$ under the state map $F^\tau$ defined above. Define
\[
    H_D=\sum_{\tau\in\cP_D}h_\tau,
    \qquad
    C_{p,\mu,D}=\norm{H_D}_{\infty,\mu}^{1/p},
\]
and assume $C_{p,\mu,D}<\infty$.
\end{assumption}

\begin{lemma}[$L_p(\mu)$ depth-$D$ contraction]
\label{lem:lp-depth}
Under \cref{ass:depth-d-coverage}, for all bounded $V,W$,
\[
    \norm{\shapley^D V-\shapley^D W}_{p,\mu}
    \le
    \gamma^D C_{p,\mu,D}\norm{V-W}_{p,\mu}.
\]
\end{lemma}

\begin{proof}
Set $E=V-W$. By \cref{lem:pointwise-depth}, for every $s\in\cS$,
\[
    \abs{(\shapley^D V)(s)-(\shapley^D W)(s)}
    \le
    \gamma^D\max_{\tau\in\cP_D}\abs{E(F^\tau(s))}.
\]
Because both sides are nonnegative and $p\ge1$, raising them to the $p$th power preserves the inequality:
\begin{align*}
    \abs{(\shapley^D V)(s)-(\shapley^D W)(s)}^p
    &\le
    \gamma^{Dp}
    \left(
        \max_{\tau\in\cP_D}\abs{E(F^\tau(s))}
    \right)^p\\
    &=
    \gamma^{Dp}
    \max_{\tau\in\cP_D}\abs{E(F^\tau(s))}^p.
\end{align*}
For any finite collection of nonnegative numbers, its maximum is no larger than its sum. Therefore,
\[
    \abs{(\shapley^D V)(s)-(\shapley^D W)(s)}^p
    \le
    \gamma^{Dp}
    \sum_{\tau\in\cP_D}\abs{E(F^\tau(s))}^p.
\]
Integrating both sides with respect to $\mu$ and interchanging the finite sum and integral gives
\begin{align*}
    \norm{\shapley^D V-\shapley^D W}_{p,\mu}^p
    &\le
    \gamma^{Dp}
    \sum_{\tau\in\cP_D}
    \int_\cS \abs{E(F^\tau(s))}^p\,d\mu(s).
\end{align*}
By the definition of a pushforward measure, each integral satisfies
\begin{align*}
    \int_\cS \abs{E(F^\tau(s))}^p\,d\mu(s)
    &=
    \int_\cS \abs{E(z)}^p\,d(F^\tau_*(\mu))(z).
\end{align*}
Absolute continuity and the definition of $h_\tau$ then imply
\begin{align*}
    \int_\cS \abs{E(z)}^p\,d(F^\tau_*(\mu))(z)
    &=
    \int_\cS \abs{E(z)}^p h_\tau(z)\,d\mu(z).
\end{align*}
Substituting these identities into the previous bound and using the definition of $H_D$ gives
\begin{align*}
    \norm{\shapley^D V-\shapley^D W}_{p,\mu}^p
    &\le
    \gamma^{Dp}
    \sum_{\tau\in\cP_D}
    \int_\cS \abs{E(z)}^p h_\tau(z)\,d\mu(z)\\
    &=
    \gamma^{Dp}
    \int_\cS \abs{E(z)}^p
    \left(\sum_{\tau\in\cP_D}h_\tau(z)\right)d\mu(z)\\
    &=
    \gamma^{Dp}
    \int_\cS \abs{E(z)}^p H_D(z)\,d\mu(z).
\end{align*}
Since $H_D(z)\le\norm{H_D}_{\infty,\mu}$ for $\mu$-almost every $z$,
\begin{align*}
    \norm{\shapley^D V-\shapley^D W}_{p,\mu}^p
    &\le
    \gamma^{Dp}\norm{H_D}_{\infty,\mu}
    \int_\cS\abs{E(z)}^p\,d\mu(z)\\
    &=
    \gamma^{Dp}C_{p,\mu,D}^p\norm{E}_{p,\mu}^p.
\end{align*}
Taking the $p$th root and recalling $E=V-W$ proves the claim.
\end{proof}

At outer iteration $k$, let $\widehat{\shapley}_{D,k}$ denote an arbitrary depth-$D$ inner-loop solver. It need not perform exact minimax backups or even be the same algorithm at every iteration. We require only a bound on its output at the current value iterate. Let $V_{k+1}$ be the fitted approximation to that output and define
\begin{align}
    \varepsilon_{\mathrm{mg},k}
    &:=
    \norm{\widehat{\shapley}_{D,k}V_k-\shapley^D V_k}_{p,\mu},
    \label{eq:mg-error}\\
    \varepsilon_{\mathrm{fit},k}
    &:=
    \norm{V_{k+1}-\widehat{\shapley}_{D,k}V_k}_{p,\mu}.
    \label{eq:fit-error}
\end{align}
\begin{lemma}[Combined one-step residual]
\label{lem:total-residual}
The inner-solver and fitting errors imply
\begin{equation}\label{eq:total-residual}
    \norm{V_{k+1}-\shapley^D V_k}_{p,\mu}
    \le
    \varepsilon_{\mathrm{mg},k}+\varepsilon_{\mathrm{fit},k}.
\end{equation}
\end{lemma}

\begin{proof}
Add and subtract the inner-solver output $\widehat{\shapley}_{D,k}V_k$ and apply the triangle inequality:
\begin{align*}
    \norm{V_{k+1}-\shapley^D V_k}_{p,\mu}
    &=
    \norm{
        V_{k+1}-\widehat{\shapley}_{D,k}V_k
        +\widehat{\shapley}_{D,k}V_k-\shapley^D V_k
    }_{p,\mu}\\
    &\le
    \norm{V_{k+1}-\widehat{\shapley}_{D,k}V_k}_{p,\mu}
    +
    \norm{\widehat{\shapley}_{D,k}V_k-\shapley^D V_k}_{p,\mu}\\
    &=
    \varepsilon_{\mathrm{fit},k}
    +\varepsilon_{\mathrm{mg},k}.
\end{align*}
This proves \cref{eq:total-residual}.
\end{proof}

\begin{lemma}[Fitted depth-$D$ minimax value iteration]
\BodyFittedDepthD
\end{lemma}

\begin{proof}
\begin{align*}
    \norm{V_{k+1}-V^*}_{p,\mu}
    &=
    \norm{V_{k+1}-\shapley^D V_k+\shapley^D V_k-V^*}_{p,\mu}\\
    &\le
    \norm{V_{k+1}-\shapley^D V_k}_{p,\mu}
    +
    \norm{\shapley^D V_k-V^*}_{p,\mu}\\
    &=
    \norm{V_{k+1}-\shapley^D V_k}_{p,\mu}
    +
    \norm{\shapley^D V_k-\shapley^D V^*}_{p,\mu} \\
    &\le \underbrace{\norm{V_{k+1}-\widehat{\shapley}_{D,k}V_k}_{p,\mu}}_{\varepsilon_{\mathrm{fit},k}}
    +
    \underbrace{\norm{\widehat{\shapley}_{D,k}V_k-\shapley^D V_k}_{p,\mu}}_{\varepsilon_{\mathrm{mg},k}}
    +
    \norm{\shapley^D V_k-\shapley^D V^*}_{p,\mu}
\end{align*}
Using the definitions of $\varepsilon_{\mathrm{mg},k}, \varepsilon_{\mathrm{fit},k}$ (\cref{lem:total-residual}) and applying \cref{lem:lp-depth} to the second term yields
\begin{align*}
    \norm{V_{k+1}-V^*}_{p,\mu}
    &\le
    \varepsilon_{\mathrm{mg},k}
    +\varepsilon_{\mathrm{fit},k}
    +\gamma^D C_{p,\mu,D}\norm{V_k-V^*}_{p,\mu}\\
    &=
    \rho\norm{V_k-V^*}_{p,\mu}
    +\varepsilon_{\mathrm{mg},k}
    +\varepsilon_{\mathrm{fit},k}.
\end{align*}
Define
\[
    e_k=\norm{V_k-V^*}_{p,\mu},
    \qquad
    \varepsilon_k
    =\varepsilon_{\mathrm{mg},k}+\varepsilon_{\mathrm{fit},k}.
\]
The preceding inequality is the scalar recursion
\begin{equation}\label{eq:scalar-recursion}
    e_{k+1}\le\rho e_k+\varepsilon_k.
\end{equation}
We can unroll this recursion as the sum
\begin{equation} \label{eq:error-sum}
    e_{K} \le
    \rho^{K}e_0
    +\sum_{j=0}^{K-1}\rho^{K-1-j}\varepsilon_j.
\end{equation}
Under the uniform error bounds,
\[
    \varepsilon_j
    \le
    \varepsilon_{\mathrm{mg}}+\varepsilon_{\mathrm{fit}}
\]
for every $j$. Substitution into \cref{eq:error-sum} gives
\begin{align*}
    e_K
    &\le
    \rho^K e_0
    +
    \left(\varepsilon_{\mathrm{mg}}+\varepsilon_{\mathrm{fit}}\right)
    \sum_{j=0}^{K-1}\rho^{K-1-j}\\
    &=
    \rho^K e_0
    +
    \left(\varepsilon_{\mathrm{mg}}+\varepsilon_{\mathrm{fit}}\right)
    \sum_{m=0}^{K-1}\rho^m\\
    &=
    \rho^K e_0
    +
    \frac{1-\rho^K}{1-\rho}
    \left(\varepsilon_{\mathrm{mg}}+\varepsilon_{\mathrm{fit}}\right), \qquad \text{(Finite geometric sum)}
\end{align*}
where the second equality uses the change of index $m=K-1-j$ and the third uses $\rho<1$. This proves \cref{eq:fitted-uniform-bound}. Finally, $\rho^K e_0\to0$ and $(1-\rho^K)/(1-\rho)\to1/(1-\rho)$, which proves the limit-superior bound.
\end{proof}

\begin{corollary}[Coverage-free fitted bound in $L_\infty$]
\label{cor:fitted-sup}
If the errors in \cref{eq:mg-error,eq:fit-error} are instead measured in $\norm{\cdot}_\infty$, then no coverage assumption is required and \cref{eq:fitted-finite-bound,eq:fitted-uniform-bound} hold with $\rho=\gamma^D$.
\end{corollary}

\begin{proof}
Repeat the proof of \cref{thm:fitted-depth-d}, replacing \cref{lem:lp-depth} with \cref{thm:depth-d-sup-contraction}.
\end{proof}

\subsection{Imperfect Regret Transfer}

\begin{definition}
\BodyRegretDefs
\end{definition}

\begin{theorem}[Imperfect regret transfer]
\BodyImperfectTransfer
\end{theorem}


\begin{proof}
With bounded utility perturbation, we can bound immediate regret perturbation:
\[
\begin{aligned}
|\reg_{i,t}'(a_i) - \reg_{i,t}(a_i)|
&=
\left|
\left[u_i'(a_i, \sigma_{-i,t}) - u_i'(\sigma_{i,t}, \sigma_{-i,t})\right]
-
\left[u_i(a_i, \sigma_{-i,t}) - u_i(\sigma_{i,t}, \sigma_{-i,t})\right]
\right| \\
&\le 2\delta_u \, .
\end{aligned}
\]
With bounded immediate regret perturbation, we can bound cumulative regret perturbation:
\[
\|\mathbf R^{c,\prime}_{i,T_1} - \mathbf R^c_{i,T_1}\|_\infty \le 2\delta_u T_1 \, .
\]
By the fitted average regret assumption and the definition $\widehat{\mathbf R}^c_{i,T_1}=T_1\widehat{\bar{\mathbf R}}_{i,T_1}$,
\[
\| \widehat{\mathbf R}^c_{i,T_1} - \mathbf R^c_{i,T_1} \|_\infty \le T_1 \delta_R \, .
\]
From Cesa-Bianchi and Lugosi~\cite[p.~10]{cesa2006prediction}:
\begin{equation} \label{eq:regret-potential-bound}
\Phi(\Racc_{i,t}) - \Phi(\Racc_{i,t-1}) \le \sum_{a_i\in\cA_i} \reg_{i,t}'(a_i)^2
\le |\cA_i|\Delta_u^2 .
\end{equation}
Therefore,
\begin{align*}
    \Phi(\Racc_{i,T}) &\le \Phi(\Racc_{i,T_1}) + T_2 |\cA_i| \Delta_u^2 \quad &\text{(\cref{eq:regret-potential-bound})} \\
    &\le wT_1|\cA_i|\Delta_u^2 + T_2|\cA_i|\Delta_u^2 \quad &\text{(\cref{eq:w-condition})}\\
    &= (wT_1 + T_2)|\cA_i|\Delta_u^2 . &
\end{align*}
The internal surrogate regret vector upon which regret matching is operating is given by
\[
\Racc_{i,T} = w\widehat{\mathbf R}^c_{i,T_1} + \sum_{t=T_1+1}^{T} \reg_{i,t}'\, ,
\]
and the true weighted regret vector in the current game that would be used by the exact non-approximated form of regret transfer~\cite{brown2014regret} is given by
\[
\mathbf S^{w,\prime}_{i,T} = w\mathbf R^{c,\prime}_{i,T_1} + \sum_{t=T_1+1}^{T} \reg_{i,t}' \, .
\]
Therefore, we can express the true weighted regret vector in the current game $\Gamma'$ as
\[
\mathbf S^{w,\prime}_{i,T} = \Racc_{i,T} + w\mathbf R^{c,\prime}_{i,T_1} - w\widehat{\mathbf R}^c_{i,T_1} \,.
\]
Because any single coordinate of a vector is bounded by the vector's $\ell_2$-norm, we can bound the internal surrogate regret as follows:
\[
\max_{a_i} \Racc_{i,T}(a_i) \le \sqrt{\Phi(\Racc_{i,T})}\le\Delta_u\sqrt{|\cA_i|(wT_1 + T_2)} \, .
\]
Following this, we can bound weighted average regret in the true new game:
\begin{align*}
    \max_{a_i}\mathbf S^{w,\prime}_{i,T}(a_i) &\le \max_{a_i}\Racc_{i,T}(a_i) + w \max_{a_i} \left(\mathbf R^{c,\prime}_{i,T_1}(a_i) - \widehat{\mathbf R}^c_{i,T_1}(a_i)\right) \\
    \frac{\max_{a_i}\mathbf S^{w,\prime}_{i,T}(a_i)}{wT_1 + T_2} &\le \frac{\Delta_u\sqrt{|\cA_i|(wT_1 + T_2)}}{wT_1 + T_2} + \frac{w\left(\|\mathbf R^{c,\prime}_{i,T_1} - \mathbf R^c_{i,T_1}\|_\infty + \|\mathbf R^c_{i,T_1} - \widehat{\mathbf R}^c_{i,T_1}\|_\infty\right)}{wT_1 + T_2} \\
    \frac{\max_{a_i}\mathbf S^{w,\prime}_{i,T}(a_i)}{wT_1 + T_2} &\le \frac{\Delta_u\sqrt{|\cA_i|}}{\sqrt{wT_1 + T_2}} + \frac{wT_1(2\delta_u +\delta_R)}{wT_1 + T_2} \, .
\end{align*}
\end{proof}

\begin{corollary}[Hannan consistency after imperfect transfer]
\BodyTransferHannan
\end{corollary}

\begin{proof}
With $T_1$, $w$, $\delta_u$, and $\delta_R$ fixed, both terms on the right-hand side of the bound in \cref{thm:imperfect-regret-transfer} vanish as $T_2\to\infty$.
\end{proof}

\begin{corollary}[Asymptotic equilibrium of averaged tree search]
\BodyAsymptoticTreeEq
\end{corollary}

\begin{proof}
By \cref{cor:transfer-hannan}, the solver without explicit exploration is Hannan consistent. Uniform $\epsilon_{\mathrm{exp}}$-exploration makes it $\epsilon_{\mathrm{exp}}$-Hannan consistent with guaranteed exploration \cite[Lemma~7]{kovavrik2015analysis}. The result then follows from \cite[Theorem~8]{kovavrik2015analysis}.
\end{proof}

\begin{lemma}[Saddle-gap stability]\label{lem:saddle-gap-stability}
Let $M$ be a two-player zero-sum payoff matrix with range at most $\Delta_u$, and write $u(x,y)=x^\top My$. If $\|\hat x-x\|_1\le\delta_1$ and $\|\hat y-y\|_1\le\delta_2$, then
\[
\max_{\tilde x}u(\tilde x,\hat y)-\min_{\tilde y}u(\hat x,\tilde y)
\le
\max_{\tilde x}u(\tilde x,y)-\min_{\tilde y}u(x,\tilde y)
+\Delta_u(\delta_1+\delta_2).
\]
\end{lemma}
\begin{proof}
Consider the payoff matrix $M$, where $M_{jk}$ is the payoff to player~1 when the players choose actions $a_{1,j}$ and $a_{2,k}$. We can then write the payoff of mixed strategies $x$ and $y$ in inner-product form as
\[
    u(x,y)=x^\top My.
\]

Because the payoff range is bounded by $\Delta_u$, we may translate all entries of $M$ by the same constant so that $M_{jk}\in[0,\Delta_u]$. This translation does not change the saddle-point gap.

With the new joint strategy $(\hat x,\hat y)$, for every mixed strategy $\tilde x$ we have
\begin{align*}
    \abs{u(\tilde x,\hat y)-u(\tilde x,y)}
    &=
    \abs{\tilde x^\top M(\hat y-y)}\\
    &=
    \abs{(\hat y-y)^\top M^\top\tilde x}\\
    &\le
    \norm{\hat y-y}_1\norm{M^\top\tilde x}_\infty \qquad\text{(Hölder's inequality)}\\
    &\le
    \delta_2\Delta_u
    .
\end{align*}

For the maximizing player,
\begin{align*}
    u(\tilde x,\hat y)
    &\le
    u(\tilde x,y)+\Delta_u\delta_2,
    \qquad \forall\,\tilde x,\\
    \max_{\tilde x}u(\tilde x,\hat y)
    &\le
    \max_{\tilde x}u(\tilde x,y)+\Delta_u\delta_2.
\end{align*}

With equivalent reasoning, for every mixed strategy $\tilde y$,
\begin{align*}
    \abs{u(\hat x,\tilde y)-u(x,\tilde y)} \le \delta_1\Delta_u, \quad
    u(\hat x,\tilde y) \ge \min_{\tilde y}u(x,\tilde y)-\Delta_u\delta_1 .
\end{align*}

Therefore,
\begin{align*}
    &\max_{\tilde x}u(\tilde x,\hat y)
    -
    \min_{\tilde y}u(\hat x,\tilde y)\\
    &\quad\le
    \max_{\tilde x}u(\tilde x,y)
    +
    \Delta_u\delta_2
    -
    \left[
        \min_{\tilde y}u(x,\tilde y)
        -
        \Delta_u\delta_1
    \right]\\
    &\quad=
    \max_{\tilde x}u(\tilde x,y)
    -
    \min_{\tilde y}u(x,\tilde y)
    +
    \Delta_u(\delta_1+\delta_2).
\end{align*}
\end{proof}

\begin{lemma}[Approximate Nash equilibrium]
\BodyApproxNash
\end{lemma}

\begin{proof}
By \cref{thm:imperfect-regret-transfer}, each player's weighted average regret in $\Gamma'$ is bounded by
\[
\mathcal R_i^{w,\prime}(T_1,T_2)
\le
\frac{\Delta_u\sqrt{|\cA_i|}}{\sqrt{wT_1 + T_2}}
+
\frac{wT_1(2\delta_u+\delta_R)}{wT_1 + T_2}.
\]
By Corollary 1 of Brown and Sandholm~\cite{brown2014regret}, if $\Gamma'$ is a two-player zero-sum game and both players use the same weighting convention, then the exact weighted average strategies form an $\epsilon_{\mathrm{reg}}$-Nash equilibrium with
\[
\epsilon_{\mathrm{reg}} = 2\max_i \mathcal R_i^{w,\prime}(T_1,T_2).
\]

Let the exact weighted average strategy be
\[
\bar\sigma_i^{w,\prime}(a)
=
\frac{
w\sum_{t=1}^{T_1}\sigma_{i,t}(a)
+
\sum_{t=T_1+1}^{T}\sigma_{i,t}(a)
}{
wT_1+T_2
}.
\]
Then
\[
\|\widehat{\bar\sigma}_i^{w,\prime}-\bar\sigma_i^{w,\prime}\|_\infty
\le
\frac{wT_1}{wT_1+T_2}\delta_\sigma,
\]
and therefore
\[
\|\widehat{\bar\sigma}_i^{w,\prime}-\bar\sigma_i^{w,\prime}\|_1
\le
|\cA_i|\frac{wT_1}{wT_1+T_2}\delta_\sigma.
\]
By \cref{lem:saddle-gap-stability}, replacing the exact weighted average strategies by the fitted weighted average strategies increases the saddle-point gap by at most
\[
\Delta_u
\left(
\|\widehat{\bar\sigma}_1^{w,\prime}-\bar\sigma_1^{w,\prime}\|_1
+
\|\widehat{\bar\sigma}_2^{w,\prime}-\bar\sigma_2^{w,\prime}\|_1
\right).
\]

Thus,
\[
\epsilon_{\mathrm{NE}}
\le
\epsilon_{\mathrm{reg}}
+
\Delta_u
\frac{wT_1}{wT_1+T_2}
\left(|\cA_1|+|\cA_2|\right)\delta_\sigma.
\]

Substituting the regret bound gives
\[
\epsilon_{\mathrm{NE}}
\le
2\max_i
\left[
\frac{\Delta_u\sqrt{|\cA_i|}}{\sqrt{wT_1 + T_2}}
+
\frac{wT_1(2\delta_u+\delta_R)}{wT_1 + T_2}
\right]
+
\Delta_u
\frac{wT_1}{wT_1+T_2}
\left(|\cA_1|+|\cA_2|\right)\delta_\sigma .
\]
\end{proof}

\subsection{End-to-end fitted-search value error}

At outer iteration $k$, let $h_s$ be the root of the depth-$D$ local game at state $s$. If $x^{h_s}_{k,\tau}$ is the player-1 payoff observation supplied to the root selection process on its $\tau$th visit, define its average observed payoff by
\[
    g^{h_s}_k(t)
    :=
    \frac{1}{t}\sum_{\tau=1}^{t}x^{h_s}_{k,\tau}.
\]
We identify the depth-$D$ approximate backup with this root statistic:
\begin{equation}\label{eq:root-average-backup}
    \bigl(\widehat{\shapley}_{D,k}^{(t)}V_k\bigr)(s)
    :=
    g^{h_s}_k(t).
\end{equation}
Thus, for a state-dependent search budget $t_k(s)$, the fitted target is
\[
    Y_k(s)
    :=
    g^{h_s}_k(t_k(s))
    =
    \bigl(\widehat{\shapley}_{D,k}^{(t_k(s))}V_k\bigr)(s).
\]
For the remainder of this subsection, set
\[
    \bigl(\widehat{\shapley}_{D,k}V_k\bigr)(s)
    :=
    \bigl(\widehat{\shapley}_{D,k}^{(t_k(s))}V_k\bigr)(s),
\]
so that this budgeted solver is the approximate operator appearing in \cref{eq:mg-error,eq:fit-error}. This is the average of the observations already supplied to the root selector, so it can be accumulated during SM-MCTS-A without separately evaluating its empirical-frequency strategy. It is the statistic denoted $g^{h_s}(t)$ in the convergence proof of Kova\v{r}\'ik and Lis\'y; it need not equal a separately stored average of raw recursive returns.

The empirical-strategy equilibrium guarantee is already
\cref{cor:transfer-sm-mcts-a}; the following lemma records only the
root-average guarantee needed for fitted value iteration.

\begin{lemma}[State-dependent-budget root-average error]
\label{lem:search-to-backup}
Suppose the range of every depth-$D$ local-game payoff is at most
$\Delta_{\mathrm{tree}}$. Run SM-MCTS-AP with uniform $\epsilon_{\mathrm{exp}}$-exploration, and fix $\eta > 0$.
Conditional on the preceding training history and $V_k$, assume for $\mu$-almost every $s$, the chosen budget satisfies $t_k(s)\ge t_{0,k}(s,\eta)$ and
\[
    \left|
        \bigl(\widehat{\shapley}_{D,k}^{(t_k(s))}V_k\bigr)(s)
        -
        \bigl(\shapley^D V_k\bigr)(s)
    \right|
    \le
    \Delta_{\mathrm{tree}}
    \left(D\epsilon_{\mathrm{exp}}+\eta\right)\,.
\]
Then, on that event,
\[
    \norm{
        \widehat{\shapley}_{D,k}V_k-\shapley^D V_k
    }_{p,\mu}
    \le
    \varepsilon_{\mathrm{mg}}
    :=
    \Delta_{\mathrm{tree}}
    \left(D\epsilon_{\mathrm{exp}}+\eta\right)\,.
\]
\end{lemma}

\begin{proof}
First normalize the local-game payoffs to an interval of width one. By
\cref{cor:transfer-hannan}, fixed finite transfer preserves Hannan
consistency. Mixing each selector with uniform exploration of probability
$\epsilon_{\mathrm{exp}}$ therefore makes it
$\epsilon_{\mathrm{exp}}$-Hannan consistent with guaranteed exploration
\cite[Lemma~7]{kovavrik2015analysis}.

Under \cref{eq:root-average-backup}, repeated application of \cite[Corollary~12]{kovavrik2015analysis} to property~1 of $(IH_D)$ from the depth-one result of \cite[Lemma~9]{kovavrik2015analysis}, followed by rescaling from unit payoff range to $\Delta_{\mathrm{tree}}$, implies that for every fixed $s$ and $\eta>0$ there is an almost-surely finite $t_{0,k}(s,\eta)$ such that
\[
    \left|
        \bigl(\widehat{\shapley}_{D,k}^{(t)}V_k\bigr)(s)
        -
        \bigl(\shapley^D V_k\bigr)(s)
    \right|
    \le
    \Delta_{\mathrm{tree}}
    \left(D\epsilon_{\mathrm{exp}}+\eta\right)
    \qquad
    \text{for every }t\ge t_{0,k}(s,\eta).
\]
The lemma assumes that this statewise conclusion holds for $\mu$-almost every $s$ and that each selected budget reaches its corresponding threshold. Therefore, on that event,
\begin{align*}
    \norm{
        \widehat{\shapley}_{D,k}V_k-\shapley^D V_k
    }_{p,\mu}
    &\le
    \norm{
        \widehat{\shapley}_{D,k}V_k-\shapley^D V_k
    }_{\infty,\mu}\\
    &\le
    \Delta_{\mathrm{tree}}
    \left(D\epsilon_{\mathrm{exp}}+\eta\right).
\end{align*}
\end{proof}

\begin{theorem}[End-to-end fitted-search value bound]
\BodyEndToEnd
\end{theorem}

\begin{proof}
For each $k$, let
\[
    \mathsf E_k
    :=
    \left\{
        \varepsilon_{\mathrm{mg},k}\le\varepsilon_{\mathrm{mg}},
        \quad
        \varepsilon_{\mathrm{fit},k}\le\varepsilon_{\mathrm{fit}}
    \right\}.
\]
The search component of $\mathsf E_k$ holds almost surely by the state-dependent-budget assumption in
\cref{lem:search-to-backup}. Applying Wilson's description-length bound
\cite{wilson2025mysterious} with confidence
$\delta/K$ gives the fitting component with conditional failure probability
at most $\delta/K$. Union bounding over $K$ iterations yields the result.
\end{proof}

\section{Algorithm Details}

\begin{algorithm}[h]
\caption{SM-MCTS-AP search at a root state}
\label{alg:search}
\BodySearch
\end{algorithm}

\begin{algorithm}[h]
\caption{Node subroutines for SM-MCTS-AP}
\label{alg:subroutines}
\textbf{\textsc{Expand}$(h)$}
\BodyExpand
\vspace{0.5em}
\textbf{\textsc{Select}$(h)$}
\BodySelect
\vspace{0.5em}
\textbf{\textsc{Update}$(h,\mathbf a,\sigma,u,v')$}
\BodyUpdate
\end{algorithm}

\subsection{Self-play training as implemented}

The main paper analyzes \mbox{SIM-AZ-$\mu$}, in which the training states at every outer iteration are drawn i.i.d.\ from a fixed distribution $\mu$. The system we actually run instead generates states by self-play, shown in \cref{alg:selfplay}.

\begin{algorithm}[h]
\caption{Self-play training for SIM-AZ (as implemented)}
\label{alg:selfplay}
\BodySelfPlay
\end{algorithm}

\end{document}